\documentclass{jfm}
\usepackage{graphicx}
\usepackage{natbib}
\usepackage{upmath}
\usepackage{amssymb}
\usepackage{amsbsy}
\usepackage{amsmath}
\usepackage{graphics}
\usepackage{subfigure}
\usepackage{dcolumn}
\usepackage{bm}
\usepackage{hyperref}
\usepackage{xcolor}
\usepackage{soul}
\usepackage{color}

\newcommand{\bc}{\begin{center}}
\newcommand{\ec}{\end{center}}

\newcommand{\be}{\begin{equation}}
\newcommand{\ee}{\end{equation}}
\newcommand{\bea}{\begin{eqnarray}}
\newcommand{\eea}{\end{eqnarray}}

\title[]{Mixing of Gaussian Solute Plumes}

\author[Daniel R. Lester]%
{Daniel R. Lester$^1$%
  \thanks{Email address for correspondence: daniel.lester@rmit.edu.au}}

\affiliation{$^1$School of Engineering, RMIT University, Melbourne, Australia}
\pubyear{2025}
\volume{}
\pagerange{}
\date{?; revised ?; accepted ?. - To be entered by editorial office}
\begin{document}

\maketitle

\begin{abstract}
We consider the general problem of mixing of Gaussian solute plumes arising across a range of different flows ranging from two-dimensional (2D) to three-dimensional (3D), steady to unsteady, regular to chaotic, Stokes to turbulent. We consider a a range of infinitesimal injection protocols (point or line sources, continuous or pulsed) which generate solute points, lines and sheets and their finite-sized counterparts (blobs, tubes and slabs). We show that evolution of this broad family of Gaussian plumes is solely governed by the initial solute distribution, the dispersion tensor and the Lagrangian history of the deformation gradient tensor. Significant simplifications arise when the deformation tensor is placed in the Protean (Lester et al, JFM, 2018) coordinate frame (which aligns with the maximum and minimum fluid stretching directions and renders the deformation tensor upper triangular), allowing evolution of the Gaussian covariance matrix to be expressed in physically relevant terms such as Lyapunov exponents, longitudinal and transverse shear rates. We couch these results in terms of evolution of the maximum concentration and scalar entropy of Gaussian plume for a range of flows and injection protocols. These results provide significant insights into the processes that govern mixing in a broad range of flows and provide the building blocks of efficient numerical methods using e.g. radial basis functions for solute transport and mixing in arbitrary flows.
\end{abstract}

\begin{keywords}
chaotic advection, stirring, dispersion, topological braiding
\end{keywords}

%%%%%%%%%%%%%%%%%%%%%%%%%%%%%%%%%%%%%%%%%%%%%%%%%%%%%%%%%%%
\section{General Framework}
\label{sec:blob}
%%%%%%%%%%%%%%%%%%%%%%%%%%%%%%%%%%%%%%%%%%%%%%%%%%%%%%%%%%%

\subsection{Evolution of a Gaussian Plume }
\label{subsec:blob}

We consider the evolution of a Gaussian plume with concentration field $C(\mathbf{x},t)$ in an unbounded (possibly porous) 2D ($d=2$) or 3D ($d=3$), steady of unsteady, compressible or incompressible fluid domain with porosity $\theta(\mathbf{x},t)$ subject to the advective velocity field $\mathbf{v}(\mathbf{x},t)$ and dispersion tensor $\mathbb{D}_0(\mathbf{x},t)$. Note for pure fluid flows, the porosity $\theta$ can be taken as unity everywhere. The initial concentration field of the blob is given by the Gaussian distribution
\begin{equation}
    C(\mathbf{x},0)=C_0(\mathbf{x})=\frac{C_0}{\sqrt{(2\pi)^n\det\boldsymbol\Sigma_0}}\exp\left(-\small{\frac{1}{2}}\mathbf{x}^\top\cdot\boldsymbol\Sigma_0^{-1}\cdot\mathbf{x}\right).\label{eqn:init}
\end{equation}
centred at the origin $\mathbf{x}=\mathbf{0}$ where $C_0$ is the mass of solute and $\boldsymbol\Sigma_0$ is the initial covariance tensor of the blob. The blob evolution is governed by the advection-dispersion equation (ADE)
\begin{equation}
\frac{\partial}{\partial t}(C\theta)+\nabla\cdot(\mathbf{v}C\theta)-\nabla\cdot(\theta\mathbb{D}_0\cdot\nabla C)=0.\label{eqn:porousADE}
\end{equation}
Under the assumption that spatial gradients of the porosity field are small compared to those of the concentration field, i.e. $|\theta\nabla C|\gg |C\nabla\theta|$, then the ADE may be expressed in terms of the volume-averaged concentration $c(\mathbf{x},t)\equiv C(\mathbf{x},t)\theta(\mathbf{x},t)$ as
\begin{equation}
\frac{\partial c}{\partial t}+\nabla\cdot(\mathbf{v}(\mathbf{x},t) c)-\nabla\cdot(\mathbb{D}_0\cdot\nabla c)=0,\label{eqn:ADE}
\end{equation}
with initial condition $c(\mathbf{x},0)=c_0(\mathbf{x})\approx C_0(\mathbf{x})\theta(\mathbf{0},0)\equiv C_0(\mathbf{x})\theta_0$. It shall prove useful to solve the ADE in the translating and rotating (denoted moving) reference frame
\begin{align}
\mathbf{x}^\prime=\mathbf{Q}(t)^\top\cdot(\mathbf{x}-\mathbf{x}_0(t)) && \mathbf{x}=\mathbf{Q}(t)\cdot\mathbf{x}^\prime+\mathbf{x}_0(t),\label{eqn:moving}
\end{align}
where $\mathbf{x}_0(t)$ is the trajectory of a fluid particle initially at the origin at time $t=0$, and $\mathbf{Q}(t)$ represents a time-dependent rotation matrix, which is a proper orthogonal transformation that satisfies
\begin{align}
\mathbf{Q}(t)^\top\cdot\mathbf{Q}(t)=\mathbf{1} & & \det[\mathbf{Q}(t)]=1,\label{eqn:proper}
\end{align}
hence $\dot{\mathbf{Q}}(t)^\top\cdot\mathbf{Q}(t)+\mathbf{Q}(t)^\top\cdot\dot{\mathbf{Q}}(t)=\mathbf{0}$. It will prove useful to determine the rotation matrix $\mathbf{Q}(t)$ in the following section, but for the present it is treated as an arbitrary matrix that satisfies (\ref{eqn:proper}). We denote the concentration field in the moving frame as
\begin{align}
	c^\prime(\mathbf{x}^\prime,t)\equiv c\left(\mathbf{Q}(t)\cdot\mathbf{x}^\prime+\mathbf{x}_0(t),t\right) && c(\mathbf{x},t)\equiv c^\prime\left(\mathbf{Q}(t)^\top\cdot(\mathbf{x}-\mathbf{x}_0(t)),t\right).\label{eqn:moving_conc} 
\end{align}
Taking the derivative of (\ref{eqn:moving}) with respect to $t$ yields the velocity in the moving frame
\begin{equation}
\begin{split}
\mathbf{v}^\prime(\mathbf{x}^\prime,t)
=&\dot{\mathbf{Q}}(t)^\top\cdot(\mathbf{x}-\mathbf{x}_0(t))+\mathbf{Q}(t)^\top\cdot(\mathbf{v}(\mathbf{x},t)-\mathbf{v}_0(t)),\\
=&\dot{\mathbf{Q}}(t)^\top\cdot\mathbf{Q}(t)\cdot\mathbf{x}^\prime+\mathbf{Q}(t)^\top\cdot(\mathbf{v}(\mathbf{x},t)-\mathbf{v}_0(t)),\label{eqn:velmove}
\end{split}
\end{equation}
where $\mathbf{v}_0(t)\equiv\mathbf{v}(\mathbf{x}_0(t),t)$. Differentiating this expression with respect to $\mathbf{x}^\prime$ then yields the moving frame velocity gradient
\begin{equation}
\nabla^\prime\mathbf{v}^\prime(\mathbf{x}^\prime,t)=\mathbf{Q}(t)^\top\cdot\nabla\mathbf{v}(\mathbf{x},t)\cdot\mathbf{Q}(t)+\dot{\mathbf{Q}}(t)^\top\cdot\mathbf{Q}(t).\label{eqn:velgradmove}
\end{equation}
Similarly, the dispersion term in (\ref{eqn:ADE}) transforms into the moving frame as
\begin{equation}
\nabla\cdot(\mathbb{D}_0(\mathbf{x},t)\cdot\nabla c)=\nabla^\prime\cdot\left(\mathbb{D}_0^\prime(t)\cdot\nabla c^\prime(\mathbf{x}^\prime,t)\right),
\end{equation}
where $\mathbb{D}_0^\prime(t)\equiv\mathbf{Q}(t)^\top\cdot\mathbb{D}_0(\mathbf{x}=\mathbf{x}_0(t),t)\cdot\mathbf{Q}(t)$, and spatial variation of the dispersion tensor around $\mathbf{x}=\mathbf{x}_0(t)$ is ignored. We show in Appendix A that this concentration field satisfies the ADE (\ref{eqn:ADE}) in the moving frame under the mappings $c\mapsto c^\prime$, $\nabla\mapsto\nabla^\prime$, $\mathbf{v}\mapsto\mathbf{v}^\prime$:
\begin{equation}
\frac{\partial c^\prime}{\partial t}+\mathbf{v}^\prime(\mathbf{x}^\prime,t)\cdot\nabla^\prime c^\prime-\nabla^\prime\cdot(\mathbb{D}^\prime_0(t)\cdot\nabla^\prime c^\prime)=0.\label{eqn:moving_ADE}
\end{equation}
The velocity field $\mathbf{v}(\mathbf{x},t)$ can be expanded to first order in space via the local velocity gradient 
\begin{equation}
\boldsymbol\epsilon(t)\equiv\nabla\mathbf{v}(\mathbf{x},t)^\top\Big|_{\mathbf{x}=\mathbf{x}_0(t)}
\end{equation}
 as
\begin{equation}
\begin{split}
\mathbf{v}(\mathbf{x},t)&=\mathbf{v}(\mathbf{x}_0(t),t)+\boldsymbol\epsilon(t)^\top\cdot(\mathbf{x}-\mathbf{x}_0(t))+\mathcal{O}(|\mathbf{x}^{\prime 2}|),\\
&=\mathbf{v}_0(t)+\boldsymbol\epsilon(t)^\top\cdot\mathbf{Q}(t)\cdot\mathbf{x}^\prime+\mathcal{O}(|\mathbf{x}^{\prime 2}|),
\end{split}
\end{equation}
Using (\ref{eqn:velmove}), the velocity field in the moving frame is expanded to first order as
\begin{equation}
\begin{split}
\mathbf{v}^\prime(\mathbf{x}^\prime,t)&=\mathbf{Q}(t)^\top\cdot\boldsymbol\epsilon(t)^\top\cdot\mathbf{Q}(t)\cdot\mathbf{x}^\prime+\dot{\mathbf{Q}}(t)^\top\cdot\mathbf{Q}(t)^\top\cdot\mathbf{x}^\prime+\mathcal{O}(|\mathbf{x}^{\prime 2}|),\\
&=\boldsymbol\epsilon^\prime(t)^\top\cdot\mathbf{x}^\prime+\mathcal{O}(|\mathbf{x}^{\prime 2}|),
\end{split}
\end{equation}
where from (\ref{eqn:velgradmove}), the velocity gradient in the moving frame is defined as
\begin{equation}
\boldsymbol\epsilon^\prime(t)\equiv\nabla^\prime\mathbf{v}^\prime(\mathbf{x}^\prime,t)^\top\Big|_{\mathbf{x}^\prime=\mathbf{0}}
\end{equation}
Hence the ADE (\ref{eqn:ADE}) in the moving coordinate frame is then
\begin{equation}
\frac{\partial}{\partial t}c^\prime(\mathbf{x}^\prime,t)+\nabla^\prime\cdot\left(\boldsymbol\epsilon^\prime(t)^\top\cdot\mathbf{x}^\prime\,c^\prime(\mathbf{x}^\prime,t)\right)-\nabla^\prime\cdot(\mathbb{D}^\prime_0(t)\cdot\nabla^\prime c^\prime)=\mathcal{O}(|\mathbf{x}^{\prime 2}|).\label{eqn:ADEmove}
\end{equation}
 
Following Risken (1996), we solve evolution of the concentration field in the moving coordinate frame by taking the Fourier transform (FT), denoted $\mathcal{F}$ of equation (\ref{eqn:ADEmove}) to yield the concentration field in frequency space as $\tilde{c}(\mathbf{k},t)\equiv\mathcal{F}[c^\prime(\mathbf{x}^\prime,t)]\equiv\int_{\mathbb{R}^d}c^\prime(\mathbf{x}^\prime,t)\exp(-i\mathbf{k}\cdot\mathbf{x}^\prime)d^n\mathbf{x}^\prime$. The temporal derivative and Laplacian operator in (\ref{eqn:ADEmove}) transform in the usual manner, and the FT of the advection term in (\ref{eqn:ADEmove}) is performed by first taking the FT of the divergence operator to yield
\begin{equation}
\mathcal{F}[\nabla^\prime\cdot(\boldsymbol\epsilon^\prime(t)^\top\cdot\mathbf{x}^\prime\,c^\prime]=i\mathbf{k}\cdot\mathcal{F}[\boldsymbol\epsilon^\prime(t)^\top\cdot\mathbf{x}^\prime\,c^\prime)],
\end{equation}
and noting that
\begin{equation}
i \boldsymbol\epsilon^\prime(t)^\top\cdot\nabla_{\mathbf{k}}\mathcal{F}[c^\prime]=\mathcal{F}[\boldsymbol\epsilon^\prime(t)^\top\cdot\mathbf{x}^\prime\,c^\prime],
\end{equation}
where $\nabla_{\mathbf{k}}$ is the gradient operator in frequency space $\mathbf{k}$. The ADE in Fourier space is then
\begin{equation}
\frac{\partial}{\partial t}\tilde{c}(\mathbf{k},t)-\boldsymbol\epsilon^\prime(t)^\top\cdot\mathbf{k}\cdot\nabla_\mathbf{k}\tilde{c}(\mathbf{k},t)+\mathbf{k}^\top\cdot\mathbb{D}_0^\prime(t)\cdot\mathbf{k}\,\tilde{c}(\mathbf{k},t)=0,\label{eqn:fourierADE}
\end{equation}

The characteristic equations for the wavenumbers $\mathbf{k}=(k_1,\dots,k_n)$
\begin{align}
&\frac{d\mathbf{k}}{dt}=-\boldsymbol\epsilon^\prime(t)^\top\cdot\mathbf{k},\quad \mathbf{k}(t=0,\mathbf{k}_0)=\mathbf{k}_0,\label{eqn:kchar}
\end{align}
simplifies (\ref{eqn:fourierADE}) to the ODE 
\begin{equation}
\frac{d\bar{c}(\mathbf{k}_0,t)}{dt}+\mathbf{k}(t;\mathbf{k}_0)^\top\cdot\mathbb{D}_0^\prime(t)\cdot\mathbf{k}(t;\mathbf{k}_0)\,\bar{c}(\mathbf{k}_0,t)=0,\label{eqn:fourierODE}
\end{equation}
in the characteristic system where $\bar{c}(\mathbf{k}_0,t)\equiv\tilde{c}(\mathbf{k}(t;\mathbf{k}_0),t)$ and $\mathbf{k}(t,\mathbf{k}_0)$ is a function of $\mathbf{k}_0$. Fourier transform of the initial condition (\ref{eqn:init}) yields $\bar{c}_0(\mathbf{k})=C_0\theta_0\exp(-\frac{1}{2}\mathbf{k}^\top\cdot\boldsymbol\Sigma_0\cdot\mathbf{k})$. Equation (\ref{eqn:fourierODE}) has the solution
\begin{equation}
\bar{c}(\mathbf{k}_0,t)=\bar{c}_0(\mathbf{k}_0)\exp\left(-\int_0^t \mathbf{k}(t^\prime;\mathbf{k}_0)^\top\cdot\mathbb{D}_0^\prime(t^\prime)\cdot\mathbf{k}(t^\prime;\mathbf{k}_0)dt^\prime\right).\label{eqn:ODEsol}
\end{equation}
From the wavenumber characteristic equation (\ref{eqn:kchar}) the wavenumbers evolve with time as
\begin{equation}
    \mathbf{k}(t;\mathbf{k}_0)=\boldsymbol\Gamma(t)\cdot\mathbf{k}_0,\quad \mathbf{k}_0=\boldsymbol\Gamma(t)^{-1}\cdot\mathbf{k}(t;\mathbf{k}_0),\label{eqn:waveevol}
\end{equation}
where $\boldsymbol\Gamma(t)$ is the fundamental solution of (\ref{eqn:kchar}) as which satisfies the ODE
\begin{align}
&\frac{d\boldsymbol\Gamma(t)}{dt}=-\boldsymbol\epsilon^\prime(t)^\top\cdot\boldsymbol\Gamma(t),\quad\boldsymbol\Gamma(t=0)=\mathbf{1}.\label{eqn:wavefund}
\end{align}
We note that the fundamental wavenumber equation (\ref{eqn:wavefund}) is very similar to the evolution equation for the deformation gradient tensor $\mathbf{F}^\prime(t;\mathbf{X}^\prime)$ that represents a linear transform between a differential material element in the Eulerian $d\mathbf{x}^\prime$ and Lagrangian $d\mathbf{X}^\prime$ frames as
\begin{equation}
d\mathbf{x}^\prime=\mathbf{F}^\prime(t,\mathbf{X}^\prime)\cdot d\mathbf{X}^\prime,
\end{equation}
where the material trajectory $\mathbf{x}^\prime(t;\mathbf{X})$ satisfies the advection equation with initial position given by the Lagrangian coordinate $\mathbf{X}^\prime$ 
\begin{equation}
\frac{d\mathbf{x}^\prime}{dt}=\mathbf{v}^\prime(\mathbf{x}^\prime(t;\mathbf{X}^\prime),t),\quad \mathbf{x}^\prime(t=0;\mathbf{X})=\mathbf{X}^\prime.\label{eqn:advect}
\end{equation}
Although the moving coordinate frame $\mathbf{x}^\prime$ is not strictly Eulerian, the linear transform (\ref{eqn:moving}) that defines this coordinate system is comprised of translations and rotations that are uniform in space (based on the trajectory $\mathbf{x}_0(t)$ that satisfies $\mathbf{x}_0(t=0)=\mathbf{0}$ and the rotation $\mathbf{Q}(t)$ respectively), hence this coordinate system does not involve material deformation and it acts as a moving Eulerian coordinate frame. Conversely, the advection equation (\ref{eqn:advect}) captures deformation of material elements with time via the mapping from moving Lagrangian $\mathbf{X}^\prime$ to moving  Eulerian $\mathbf{x}^\prime$ coordinates. We note that by definition, the origin $\mathbf{x}=\mathbf{0}$ is invariant in the moving coordinate frame, hence the Eulerian and Lagrangian frames coincide at this point $\mathbf{x}^\prime(t;\mathbf{X}=\mathbf{0})=\mathbf{0}$ for all $t$, but this property does not hold in general. As the solute blob is always centred at the origin $\mathbf{x}^\prime=\mathbf{0}$ of the moving Eulerian frame, we are primarily interested in fluid deformation local to this point $\mathbf{x}^\prime=\mathbf{X}^\prime=\mathbf{0}$, which is quantified by the deformation tensor denoted $\mathbf{F}^\prime(t)\equiv \mathbf{F}^\prime(t;\mathbf{X}=\mathbf{0})$. The evolution equation for $\mathbf{F}^\prime(t)$ is
\begin{align}
&\frac{d\mathbf{F}^\prime(t)}{dt}=\boldsymbol\epsilon^\prime(t)\cdot\mathbf{F}^\prime(t),\quad\mathbf{F}^\prime(t=0)=\mathbf{1}.\label{eqn:deform}
\end{align}
Comparison with (\ref{eqn:wavefund}) shows that the fundamental solution of the wavenumber characteristic equation is equal to the inverse of the deformation gradient tensor:
\begin{equation}
\boldsymbol\Gamma(t)=\mathbf{F}^\prime(t)^{-1},
\end{equation}
hence the wavenumbers in (\ref{eqn:waveevol}) evolve with fluid deformation as
\begin{equation}
    \mathbf{k}(t;\mathbf{k}_0)=\mathbf{F}^\prime(t)^{-1}\cdot\mathbf{k}_0,\quad \mathbf{k}_0=\mathbf{F}^\prime(t)\cdot\mathbf{k}(t;\mathbf{k}_0).
\end{equation}
Substitution of these relations into (\ref{eqn:ODEsol}) then yields 
\begin{equation}
\bar{c}(\mathbf{k}_0,t)=\bar{c}_0(\mathbf{k}_0)\exp\left(-\mathbf{k}_0^\top\cdot\int_0^t \mathbf{F}^\prime(t^\prime)^{-1}\cdot\mathbb{D}_0^\prime(t^\prime)\cdot\mathbf{F}^\prime(t^\prime)^{-\top} dt^\prime\cdot\mathbf{k}_0\right).\label{eqn:ODEsol}
\end{equation}
and conversion to $\tilde{c}(\mathbf{k},t)$ yields
\begin{equation}
\tilde{c}(\mathbf{k},t)=c_0\exp\left(-\frac{1}{2}\mathbf{k}^\top\cdot\boldsymbol\Sigma(t)\cdot\mathbf{k}\right),\label{eqn:ADEsolnF}
\end{equation}
where the covariance matrix evolves as
\begin{equation}
\boldsymbol\Sigma(t)\equiv\boldsymbol\Delta(t)+2\boldsymbol\Lambda(t),\quad \boldsymbol\Sigma(t=0)=\boldsymbol\Sigma_0.
\end{equation}
Here $\boldsymbol\Delta(t)$ encodes reversible deformation of the initial blob $c_0(\mathbf{x})$ due to fluid deformation
\begin{align}
\boldsymbol\Delta(t)=\mathbf{F}^\prime(t)\cdot\boldsymbol\Sigma_0\cdot\mathbf{F}^\prime(t)^\top,\label{eqn:stretchhistory}
\end{align}
and $\boldsymbol\Lambda(t)$ represents irreversible variance growth of the blob due to augmented diffusion via the deformation history along the fluid trajectory $\mathbf{x}_0(t)$:
\begin{align}
&\boldsymbol\Lambda(t)\equiv\mathbf{F}^\prime(t)\cdot\left(\int_0^t \mathbf{F}^\prime(t^\prime)^{-1}\cdot\mathbb{D}_0^\prime(t^\prime)\cdot\mathbf{F}^\prime(t^\prime)^{-\top}dt^\prime\right)\cdot\mathbf{F}^\prime(t)^\top.
\end{align}
Taking the inverse Fourier transform of (\ref{eqn:ADEsolnF}) then yields the blob concentration field in the moving frame as
\begin{equation}
    c^\prime(\mathbf{x}^\prime,t)=\frac{c_0}{\sqrt{(2\pi)^n\det\boldsymbol\Sigma(t)}}\exp\left(-\frac{1}{2}\mathbf{x}^{\prime\top}\cdot\boldsymbol\Sigma^{-1}(t)\cdot\mathbf{x}^\prime\right),\label{eqn:soln}
\end{equation}
where $c(\mathbf{x},t)=c^\prime(\mathbf{Q}(t)^\top\cdot(\mathbf{x}-\mathbf{x}_0(t)),t)$. The maximum concentration of the blob then evolves as
\begin{equation}
c_m(t)=c^\prime(\mathbf{0},t)=\frac{c_0}{\sqrt{(2\pi)^n\det\boldsymbol\Sigma(t)}}.
\end{equation}

Via the reversible $\boldsymbol\Delta(t)$ and irreversible $\boldsymbol\Lambda(t)$ contributions to the covariance matrix, evolution of the solute blob is controlled by the evolution of fluid deformation gradient tensor $\mathbf{F}^\prime(t)$ in the moving coordinate frame. In absence of fluid deformation (i.e. in a uniformly translating and rotating velocity field), the velocity gradient tensor $\boldsymbol\epsilon^\prime(t)$ is zero, and the deformation tensor is identity ($\mathbf{F}^\prime(t)=\mathbf{1}$). In this instance, the covariance matrix grows as a simple dispersion process: $\boldsymbol\Sigma(t)=\boldsymbol\Sigma_0+2 t\mathbb{D}_0^\prime(t)$.

Conversely, in the presence of fluid deformation (stretching, shearing, rotation), the elements of $\mathbf{F}^\prime(t)$ can evolve exponentially with time, leading to significantly augmented dispersion. Expression of the covariance matrix as
\begin{equation}
\boldsymbol\Sigma(t)=\mathbf{F}^\prime(t)\cdot\left(\boldsymbol\Sigma_0+2\int_0^t \mathbf{F}^\prime(t^\prime)^{-1}\cdot\mathbb{D}_0^\prime(t^\prime)\cdot\mathbf{F}^\prime(t^\prime)^{-\top}dt^\prime\right)\cdot\mathbf{F}^\prime(t)^\top\label{eqn:altcovar}
\end{equation}
highlights that the history of fluid compression (quantified by $\int_0^t\mathbf{F}^\prime(t^\prime)^{-1}\cdot\mathbf{F}^\prime(t^\prime)^{-\top}dt^\prime$ in (\ref{eqn:altcovar})) accelerates solute diffusion relative the fluid via the sharpening of spatial concentration gradients. Instantaneous fluid stretching (quantified by $\mathbf{F}^\prime(t)\cdot(\dots)\cdot\mathbf{F}^\prime(t)^\top$ in (\ref{eqn:altcovar})) acts to alter concentration variance via deformation of the solute blob via deformation of the supporting fluid, but does not drive diffusion of the solute relative to the fluid. As such, the compression history controls dilution of the blob into the surrounding fluid, whereas the instantaneous stretching controls the blob dispersion. For simplicity of exposition we hereafter only consider divergence-free flows (i.e. $\det(\mathbf{F})=1$) and isotropic dispersion ($\mathbb{D}_0=D_m\mathbf{I}$) but note that these variations can easily be included in the final formulation.

This augmentation of dispersion is illustrated by considering oscillatory flows with zero net fluid displacement. For simplicity of exposition, if one considers a time periodic flow (with period $T$, such that $\mathbf{v}(\mathbf{x},t+nT)=\mathbf{v}(\mathbf{x},t)$ for any integer $n$) with zero net displacement over a period (such that $\mathbf{x}(nT;\mathbf{X})=\mathbf{X}$ for all $\mathbf{X}$ and $\mathbf{F}(nT)=\mathbf{1}$), then the covariance matrix at integer multiples of the flow period is $\boldsymbol\Sigma(nT)=\boldsymbol\Sigma_0+2D_mn\mathbf{M}(T)$, where is the accelerated dispersivity over a period due to fluid deformation.

There exist several measures of mixing that reflect the production of entropy due to dissipation of the concentration field. One such measure is the dilution index (Kitanidis 1994) $E(t)=\exp(H(t))$, which is related to the entropy of the concentration field as
\begin{equation}
H(t)=-\int_{\mathbb{R}^d} c(\mathbf{x},t)\ln[c(\mathbf{x},t)]d^d\mathbf{x}.
\end{equation}
For a Gaussian concentration field, this is related to the maximum concentration as
\begin{equation}
E(t)=\frac{e}{c_m(t)}=\frac{e}{c_0}\sqrt{(2\pi)^d\det\boldsymbol\Sigma(t)}.
\end{equation}
For a point-wise initial blob (where $\boldsymbol\Sigma_0=\mathbf{0}$) with $\det\mathbf{F}=1$, $\det\boldsymbol\Sigma(t)$ simplifies to $(2D_m)^d\det\mathbf{M}(t)$, where
\begin{equation}
\mathbf{M}(t)\equiv\int_0^t \mathbf{C}(t)^{-1}dt^\prime,
\end{equation}
where $\mathbf{C}(t)\equiv\mathbf{F}^\prime(t^\prime)^{\top}\cdot\mathbf{F}^\prime(t^\prime)$ is the right Cauchy-Green tensor, quantifies the accumulated deformation of the material element over time, and so the dilution index evolves as
\begin{equation}
E(t)=\frac{e}{c_0}\sqrt{(4\pi D_m)^d\det\mathbf{M}(t)},
\end{equation}
hence dilution of the blob is controlled entirely by $\det{\mathbf{M}(t)}$. In the following Section we develop simplified expressions for $\det\mathbf{M}(t)$ for solute blobs, lines and sheets.

%%%%%%%%%%%%%%%%%%%%%%%%%%%%%%%%%%%%%%%%%%%%%%%
\section{Plume Evolution in Protean Coordinates}
%%%%%%%%%%%%%%%%%%%%%%%%%%%%%%%%%%%%%%%%%%%%%%%

In the previous section, the rotation operator $\mathbf{Q}(t)$ central to the moving coordinate frame $\mathbf{x}^\prime$ was deliberately assigned to be arbitrary. Here we show that a judicious choice of $\mathbf{Q}(t)$ simplifies calculation of the deformation gradient tensor $\mathbf{F}^\prime(t)$ and provides insights into how the fluid kinematics govern deformation and thus blob dilution and dispersion. Various classes of flows - 2D/3D, steady/unsteady, chaotic/non-chaotic - exhibit markedly different dynamics with respect to evolution of the deformation tensor $\mathbf{F}^\prime(t)$. In general, chaotic flows (in the Lagrangian sense) are characterised by exponential stretching of material elements and thus exponential asymptotic growth of $|\mathbf{F}^\prime(t)|$, whereas non-chaotic flows exhibit power-law asymptotic growth of $|\mathbf{F}^\prime(t)|$. The demarcation of chaotic and non-chaotic flows may not be immediately obvious, but is dictated by constraints on the Lagrangian kinematics of these flows.

The Protean coordinate frame~\citep{Lester:2018aa} provides a natural framework that ensures these constraints are enforced and provides a direct physical interpretation of the elements of the the velocity gradient tensor, as well as greatly simplifying computation of fluid deformation. The deformation tensor in the moving and stationary coordinate frames are related as
\begin{equation}
\mathbf{F}^\prime (t)=\mathbf{Q}^\top(t)\cdot\mathbf{F}(t)\cdot\mathbf{Q}(0).
\end{equation}
The main concept behind the Protean coordinate system is that regardless of whether the flow is steady or unsteady, 2D or 3D, chaotic or non-chaotic, the rotation operator $\mathbf{Q}(t)$ is chosen such that the resultant velocity gradient tensor $\boldsymbol\epsilon^\prime(t)$ in the moving coordinate frame (\ref{eqn:moving}) is upper triangular, such that for 2D and 3D flows
\begin{align}
\boldsymbol\epsilon_{2\text{D}}^\prime(t)=\left(
\begin{array}{cc}
\epsilon^\prime_{11}(t) & \epsilon^\prime_{12}(t)\\
0 & \epsilon^\prime_{22}(t)
\end{array}\right)
&&
\boldsymbol\epsilon_{3\text{D}}^\prime(t)=\left(
\begin{array}{ccc}
\epsilon^\prime_{11}(t) & \epsilon^\prime_{12}(t) & \epsilon^\prime_{13}(t)\\
0 & \epsilon^\prime_{22}(t) & \epsilon^\prime_{23}(t)\\
0 & 0 & \epsilon^\prime_{33}(t),
\end{array}\right)
\label{eqn:epsilon}
\end{align}
where $\epsilon^\prime_{ii}(t)$ quantifies fluid stretching and $\epsilon^\prime_{ij}(t), j>i$ represent shear deformation and vorticity. The ensemble averages $\langle\epsilon^\prime_{ii}\rangle$ denote the Lyapunov exponents $\lambda_i$ of the system, which are constrained according to
\begin{equation}
\sum_{i=1}^d\lambda_i=0,\label{eqn:lyapzero}
\end{equation}
even for non divergence-free flows. For incompressible flows, the divergence-free condition imposes the additional constraint $\sum_{i=1}^d\epsilon_{ii}(t)=0$. As the velocity gradient $\boldsymbol\epsilon^\prime(t)$ in the Protean frame is upper triangular, from (\ref{eqn:deform}) the deformation gradient tensor $\mathbf{F}^\prime(t)$ is also upper triangular, and the components (whether 2D or 3D) have the explicit solutions
\begin{align}
&F_{ij}^\prime(t)=0,\quad i<j\\
&F_{ii}^\prime(t)=\exp\left(\int_0^t\epsilon^\prime_{ii}(t^\prime)dt^\prime\right),\quad i=1:d,\label{eqn:Fdiag}\\
&F_{12}^\prime(t)=F^\prime_{11}(t)\int_0^t\frac{\epsilon^\prime_{12}(t^\prime)F^\prime_{22}(t^\prime)}{F^\prime_{11}(t^\prime)}dt^\prime,\label{eqn:F12}\\
&F_{23}^\prime(t)=F^\prime_{22}(t)\int_0^t\frac{\epsilon^\prime_{23}(t^\prime)F^\prime_{33}(t^\prime)}{F^\prime_{22}(t^\prime)}dt^\prime,\label{eqn:F23}\\
&F_{13}^\prime(t)=F^\prime_{11}(t)\int_0^t\frac{\epsilon^\prime_{12}(t^\prime)F^\prime_{23}(t^\prime)+\epsilon^\prime_{13}(t^\prime)F^\prime_{33}(t^\prime)}{F^\prime_{11}(t^\prime)}dt^\prime.\label{eqn:F13}
\end{align}
As we will exclusively use the Protean frame throughout the remainder of this paper, we shall henceforth drop the primes associated with this frame.
\iffalse
From (), the inverse of the right Cauchy-Green tensor is then
\begin{equation}
\mathbf{C}(t)^{-1}=\left(
\begin{array}{ccc}
 \frac{F_{12}^\prime(t)^2}{F_{11}^\prime(t)^2
   F_{22}^\prime(t)^2}+\frac{1}{F_{11}^\prime(t)^2}+\frac{\left(F_{12}^\prime(t)
   F_{23}^\prime(t)-F_{13}^\prime(t) F_{22}^\prime(t)\right){}^2}{F_{11}^\prime(t)^2
   F_{22}^\prime(t)^2 F_{33}^\prime(t)^2} & -\frac{F_{12}^\prime(t)}{F_{11}^\prime(t)
   F_{22}^\prime(t)^2}-\frac{F_{23}^\prime(t) \left(F_{12}^\prime(t)
   F_{23}^\prime(t)-F_{13}^\prime(t) F_{22}^\prime(t)\right)}{F_{11}^\prime(t) F_{22}^\prime(t)^2
   F_{33}^\prime(t)^2} & \frac{F_{12}^\prime(t) F_{23}^\prime(t)-F_{13}^\prime(t)
   F_{22}^\prime(t)}{F_{11}^\prime(t) F_{22}^\prime(t) F_{33}^\prime(t)^2} \\
 -\frac{F_{12}^\prime(t)}{F_{11}^\prime(t) F_{22}^\prime(t)^2}-\frac{F_{23}^\prime(t)
   \left(F_{12}^\prime(t) F_{23}^\prime(t)-F_{13}^\prime(t)
   F_{22}^\prime(t)\right)}{F_{11}^\prime(t) F_{22}^\prime(t)^2 F_{33}^\prime(t)^2} &
   \frac{F_{23}^\prime(t)^2}{F_{22}^\prime(t)^2
   F_{33}^\prime(t)^2}+\frac{1}{F_{22}^\prime(t)^2} &
   -\frac{F_{23}^\prime(t)}{F_{22}^\prime(t) F_{33}^\prime(t)^2} \\
 \frac{F_{12}^\prime(t) F_{23}^\prime(t)-F_{13}^\prime(t) F_{22}^\prime(t)}{F_{11}^\prime(t) F_{22}^\prime(t)
   F_{33}^\prime(t)^2} & -\frac{F_{23}^\prime(t)}{F_{22}^\prime(t) F_{33}^\prime(t)^2} &
   \frac{1}{F_{33}^\prime(t)^2} \\
\end{array}
\right)
\end{equation}
\fi

%%%%%%%%%%%%%%%%%%%%%%%%%%%%%%%%%%%%%%%%%%%%%%%%%%%
\section{Points, Lines and Sheets}

Different solute injection protocols such as line or point sources, continuous or pulsed injection generate different solute distributions that can be compactly represented in the Protean frame. These different injection protocols can be handled via the Green's function
\begin{equation}
G(\mathbf{x},t|\mathbf{x}',t')=\frac{\exp\left[-\frac{1}{2}(\mathbf{x}-\boldsymbol\Phi(t;\mathbf{x}',t'))^\top\cdot\boldsymbol\Sigma(t;\mathbf{x}',t')^{-1}\cdot(\mathbf{x}-\boldsymbol\Phi(t;\mathbf{x}',t'))\right]}
{\sqrt{(2\pi)^2\det[\boldsymbol\Sigma(t;\mathbf{x}',t')]}},
\end{equation}
for the transport equation
\begin{equation}
\frac{\partial c}{\partial t}+\nabla\cdot(\mathbf{v}(\mathbf{x},t) c)-\nabla\cdot(\mathbb{D}_0\cdot\nabla c)=\delta(t-t')\delta(\mathbf{x}-\mathbf{x}'),\label{eqn:greenADE}
\end{equation}
where $\boldsymbol\Phi(t;\mathbf{x}',t')$ is the position of a tracer particle at time $t$ which was initially at position $\mathbf{x}'$ at time $t'$, which is given by the advection equation as
\begin{equation}
\frac{\partial\boldsymbol\Phi(t;\mathbf{x}',t')}{\partial t}=\mathbf{v}\left[\boldsymbol\Phi(t;\mathbf{x}',t'),t\right],\quad \boldsymbol\Phi(t';\mathbf{x}',t')=\mathbf{x}',
\end{equation}
and the covariance tensor $\boldsymbol\Sigma(t;\mathbf{x}',t')$ is also given as
\begin{equation}
\boldsymbol\Sigma(t;\mathbf{x}',t')=\mathbf{F}^\prime(t;\mathbf{x}',t')\cdot\left(2\int_0^t d\tau\,\mathbf{F}^\prime(\tau;\mathbf{x}',t')^{-1}\cdot\mathbb{D}_0^\prime(\tau)\cdot\mathbf{F}^\prime(\tau;\mathbf{x}',t')^{-\top}\right)\cdot\mathbf{F}^\prime(t;\mathbf{x}',t')^\top,\label{eqn:Sigma_green}
\end{equation}
where the Protean deformation tensor $\mathbf{F}$ satisfies
\begin{align}
&\frac{d\mathbf{F}^\prime(t;\mathbf{x}',t')}{dt}=\boldsymbol\epsilon^\prime(t;\mathbf{x}',t)\cdot\mathbf{F}^\prime(t;\mathbf{x}',t'),\quad\mathbf{F}^\prime(t';\mathbf{x}',t')=\mathbf{1},\label{eqn:deform_Green}
\end{align}
with
\begin{equation}
\boldsymbol\epsilon^\prime(t;\mathbf{x}',t)\equiv[\nabla'\mathbf{v}'(\mathbf{x},t)]^\top\Big|_{\mathbf{x}=\boldsymbol\Phi(t;\mathbf{x}',t')}.
\end{equation}
Note that for steady flows, functions (such as $\boldsymbol\Phi$, $\mathbf{F}^\prime$, $\boldsymbol\Sigma$) of the form $R(t;\mathbf{x}',t')\mapsto R(t-t';\mathbf{x}')$, and so the Green's function also simplifies as
\begin{equation}
G(\mathbf{x},t|\mathbf{x}',t')\mapsto G(t-t',\mathbf{x}|\mathbf{x}').
\end{equation}
In general, the concentration field arising from the source term $S(\mathbf{x},t)$ is then
\begin{equation}
c(\mathbf{x},t)=\int_\Omega\int_{-\infty}^t G(\mathbf{x},t|\mathbf{x}',t') S(\mathbf{x}',t')dt'\,d^n\mathbf{x}'.\label{eqn:conc_Green}
\end{equation}
Note that for a pulsed Gaussian source term of the form
\begin{equation}
S(\mathbf{x},t)=\delta(t)\frac{\exp\left(-\frac{1}{2}(\mathbf{x}-\mathbf{x}_0)^\top\cdot\boldsymbol\Sigma_0^{-1}\cdot(\mathbf{x}-\mathbf{x}_0)\right)}
{\sqrt{(2\pi)^2\det[\boldsymbol\Sigma_0]}},
\end{equation}
(\ref{eqn:conc_Green}) recovers the expression (\ref{eqn:soln}) for $c$ with $\boldsymbol\Sigma(t)$ given by (\ref{eqn:altcovar}). 

\subsection{Continuous Injection in Steady Flow}\label{subsec:continuous}

For a continuous injection protocol with source term $S(\mathbf{x},t)=\delta(\mathbf{x}-\mathbf{x}')H(t)$ (with $H(t)$ the Heaviside step function), in a steady flow the concentration field is given by
\begin{equation}
c(\mathbf{x},t)=\int_{0}^t G(t-t',\mathbf{x}|\mathbf{x}') dt',\label{eqn:continuous_transient}
\end{equation}
which converges to the steady plume given by the Laplace transform $G^*$ of the Green's function as
\begin{equation}
c_\infty(\mathbf{x})\equiv\lim_{t\rightarrow\infty}c(\mathbf{x},t)=G^*(s,\mathbf{x}|\mathbf{x}')|_{s=0}\equiv\int_0^\infty\exp(-st)G(t-t',\mathbf{x}|\mathbf{x}') dt'|_{s=0}.\label{eqn:continuous_steady}
\end{equation}
For many flows, this integral is non-trivial to compute, hence an alternate approach is to assume concentration gradients under continuous injection are negligible in the $x_1$-direction in the Protean frame, leading to the augmented ADE
\begin{equation}
v_1(\mathbf{x})\frac{\partial c_\infty(\mathbf{x})}{\partial x_1}+\nabla_\bot\cdot(\mathbf{v}_\bot(\mathbf{x}) c_\infty(\mathbf{x}))-\nabla_\bot\cdot(\mathbb{D}_0\cdot\nabla_\bot c_\infty(\mathbf{x}))=\delta(\mathbf{x}-\mathbf{x}'),\label{eqn:steady_ADE}
\end{equation}
where $\mathbf{a}_\bot$ denotes the transverse components of $\mathbf{a}$, i.e. $\mathbf{a}_\bot=(a_2,a_3)$. Introducing the Lagrangian travel time $\tau$, such that
\begin{equation}
\frac{\partial x_1(\tau;x_2,x_3)}{\partial \tau}=v_1(x_1(\tau;x_2,x_3),x_2,x_3),\quad x_1(\tau=0,x_2,x_3)=0,
\end{equation}
the coordinate system is mapped as $\mathbf{x}=(x_1,x_2,x_3)\mapsto(\tau,x_2,x_3)=(\tau,\mathbf{x}_\bot)$, this ADE may be written as
\begin{equation}
\frac{\partial c_\infty(\tau,\mathbf{x}_\bot)}{\partial\tau}+\nabla_\bot\cdot(\mathbf{v}_\bot(\mathbf{x}_\bot,\tau) c_\infty(\tau,\mathbf{x}_\bot))-\nabla_\bot\cdot(\mathbb{D}_0\cdot\nabla_\bot c_\infty(\tau,\mathbf{x}_\bot))=\delta(\mathbf{x}_\bot-\mathbf{x}^\prime_\bot)\delta(\tau-\tau'),\label{eqn:2D_ADE}
\end{equation}
which is a 2D analogue of (\ref{eqn:ADE}), and so has the continuous Green's function $G_\bot\approx G^*$, where
\begin{equation}
G_\bot(\tau-\tau',\mathbf{x}_\bot|\mathbf{x}^\prime_\bot)=\frac{\exp\left[-\frac{1}{2}(\mathbf{x}_\bot-\boldsymbol\Phi_\bot(\tau-\tau';\mathbf{x}^\prime_\bot))^\top\cdot\boldsymbol\Sigma_\bot(\tau-\tau';\mathbf{x}^\prime_\bot)^{-1}\cdot(\mathbf{x}_\bot-\boldsymbol\Phi_\bot(\tau-\tau';\mathbf{x}^\prime_\bot))\right]}
{\sqrt{(2\pi)^2\det[\boldsymbol\Sigma_\bot(\tau-\tau';\mathbf{x}^\prime_\bot)]}},
\end{equation}
where $\boldsymbol\Sigma_\bot(\tau-\tau';\mathbf{x}^\prime_\bot)$ is the following submatrix of $\boldsymbol\Sigma$
\begin{equation}
\boldsymbol\Sigma_\bot(\tau-\tau';\mathbf{x}^\prime_\bot)\equiv
\left(\begin{array}{cc}
\Sigma_{22} & \Sigma_{23}\\
\Sigma_{23} & \Sigma_{33}
\end{array}\right).
\end{equation}
%and the Lagrangian travel times $t$, $t'$ respectively serve as proxies for $x_1$, $x_1'$ via the mapping $x_1(t;x_2,x_3)$, where
%\begin{equation}
%\frac{\partial x_1(t;x_2,x_3)}{\partial t}=v(x_1(t;x_2,x_3),x_2,x_3),\quad x_1(0;x_2,x_3)=0.
%\end{equation}

\subsection{Point Sources}

The results can be used to derive results for mixing of a continuously injected and pulsed point source. For the continuous case, the source term is $S(\mathbf{x},t)=\delta(\mathbf{x})$, hence $c_\infty(\mathbf{x})=G_\bot(\tau;\mathbf{x}|\mathbf{0})$ and so
\begin{equation}
c_\infty(\tau,\mathbf{x}_\bot)=\frac{\exp\left[-\frac{1}{2}(\mathbf{x}_\bot-\mathbf{x}_{0,\bot}(\tau))^\top\cdot\boldsymbol\Sigma_\bot(\tau)^{-1}\cdot(\mathbf{x}_\bot-\mathbf{x}_{0,\bot}(\tau))\right]}
{\sqrt{(2\pi)^{d=2}\det[\boldsymbol\Sigma_\bot(\tau)]}},\label{eqn:conc_cts_point}
\end{equation}
where $\mathbf{x}_0(\tau)_\bot\equiv\boldsymbol\Phi_\bot(\tau;\mathbf{0})$, and $\boldsymbol\Sigma_\bot(\tau)\equiv\boldsymbol\Sigma_\bot(\tau;\mathbf{0})$. Hence the dilution index and maximum concentration evolve as
\begin{equation}
E(\tau)=\frac{e}{c_m(\tau)}=\frac{e}{c_0}\sqrt{(2\pi)^{d=2}\det\boldsymbol\Sigma_\bot(\tau)},\quad \det\boldsymbol\Sigma_\bot(t)=\frac{M'_{22}(t)M'_{33}(t)-M'_{23}(t)^2}{F'_{11}(t)^2}.
\end{equation}
Conversely, for a continuous point-wise injection, the source term is $S(\mathbf{x},t)=\delta(\mathbf{x})\delta(t)$, hence $c(\mathbf{x},t)=G(t;\mathbf{x}|\mathbf{0})$ and so $c(\mathbf{x},t)$ is given by (\ref{eqn:conc_cts_point}) with $d\mapsto 3$, $\tau\mapsto t$, $\mathbf{x}_\bot\mapsto\mathbf{x}$, $\mathbf{x}_{0,\bot}(\tau)\mapsto\mathbf{x}_0(t)$ and $\boldsymbol\Sigma_\bot(\tau)\mapsto\boldsymbol\Sigma(t)$. Hence the dilution index and maximum concentration evolve with the Lagrangian travel time $\tau$ as
\begin{equation}
\begin{split}
E(\tau)&=\frac{e}{c_m(\tau)}=\frac{e}{c_0}\sqrt{(2\pi)^{d=3}\det\boldsymbol\Sigma(\tau)},\\
\det\boldsymbol\Sigma(t)&=M_{11} M_{22} M_{33}+2 M_{13} M_{23} M_{12}-M_{11}
   M_{23}^2-M_{13}^2 M_{22}-M_{33} M_{12}^2,\\
   &=M_{11}F_{11}^2\det\boldsymbol\Sigma_\bot(t)+2 M_{12} M_{13} M_{23}-M_{13}^2 M_{22}-M_{12}^2 M_{33}. 
   \end{split}
\end{equation}
Hence the longitudinal shears quantified by $M_{12}$, $M_{13}$ account for the additional dilution experienced by the pulsed injection.

\subsection{Line Sources}

We also consider mixing due to a continuously injected line source along the stretching direction $x_2$ in the Protean frame, given by the source term $S(\mathbf{x},t)=\delta(x_3)$, such that a non-diffusive solute forms a 2D sheet spanned by the $x_1$ and $x_2$ coordinates. For this injection protocol, the steady concentration field is given by
\begin{equation}
c_\infty(\mathbf{x}_\bot,\tau)=\int_{-\infty}^\infty G_\bot(\tau,x_2,x_3|x_2^\prime,0)dx_2,
\end{equation}
which is in general is non-trivial to evaluate. Similar to the continuous injection Green's function, we invoke the assumption that gradients in the $x_2$ are negligible, hence the ADE simplifies to
\begin{equation}
\frac{\partial c_\infty(\tau,\mathbf{x}_\bot)}{\partial\tau}+v_3(\mathbf{x})\frac{\partial c_\infty(\mathbf{x})}{\partial x_3}-\frac{\partial}{\partial x_3}\left(D_{33}\frac{\partial c}{\partial x_3}\right)=\delta(x_2-x_2')\delta(\tau-\tau'),\label{eqn:1D_ADE}
\end{equation}

\subsection{Sheets}\label{subsec:sheets}

Continuous injection of a non-diffusive line source of initial width $\Sigma_{0,33}(X_2)$ (where $X_2$ is the $x_2$ Lagrangian coordinate) oriented along the $x_2$-direction in the Protean frame generates a sheet in the $x_1-x_2$ plane of the Protean frame. As such, for diffusive solutes there are negligible concentration gradients in both the $x_1$-direction and the $x_2$-directions, hence only the 33-component of the covariance matrix $\boldsymbol\Sigma(t)$ plays a role, where
\begin{equation}
\Sigma_{33}(t,X_2)=F^\prime_{3,3}(t,X_2)^2\left(\Sigma_{0,33}(x_2)+2D_m\int_0^t F_{33}^\prime(t^\prime,X_2)^{-2}dt^\prime\right),
\label{eqn:sigma_33}
\end{equation}
which for chaotic flows with $F_{33}^\prime(t,X_2)\approx\exp(-\lambda_\infty(X_2) t)$ recovers the classic Batchelor scale $s_B$ as
\begin{equation}
\Sigma_{33}(t,X_2)\approx e^{-2\lambda_\infty(X_2) t}\left(\Sigma_{0,33}(X_2)+\frac{D_m(e^{2\lambda_\infty(X_2) t}-1)}{\lambda_\infty(X_2)}+\right)\rightarrow\frac{D_m}{\lambda_\infty(X_2)}=s_B(X_2)^2.
\label{eqn:sigma_33_sB}
\end{equation}

\subsection{Pulsed Lines}

Let us consider pulsed injection of line source in a 2D flow with steady velocity field $\mathbf{v}=(\lambda_\infty x_1,-\lambda_\infty x_2)$, such that the deformation and covariance tensors are
\begin{equation}
\mathbf{F}(t)=\left(\begin{array}{cc}
\exp(\lambda_\infty t) & 0 \\
0 & \exp(-\lambda_\infty t)
\end{array}\right),\quad
\boldsymbol\Sigma(t)=\left(\begin{array}{cc}
\frac{D_m}{\lambda_\infty}\left(\exp(2\lambda_\infty t)-1\right) & 0 \\
0 & \frac{D_m}{\lambda_\infty}\left(1-\exp(-2\lambda_\infty t)\right)
\end{array}\right).
\end{equation}
Hence the Green's function for this flow is
\begin{equation}
G(\mathbf{x},t|\mathbf{x}',t')=\frac{1}{\sqrt{(2\pi)^2\det(\boldsymbol\Sigma(t-t'))}}\exp\left(-\frac{1}{2}(\mathbf{x}-\mathbf{x}')^\top\cdot\boldsymbol\Sigma(t-t')^{-1}\cdot(\mathbf{x}-\mathbf{x}')\right),
\end{equation}
and the resultant concentration field arising from the source term $S$ is, ingeneral
\begin{equation}
c(\mathbf{x},t)=\int_\Omega\int_{-\infty}^t G(\mathbf{x},t|\mathbf{x}',t') S(\mathbf{x}',t')dt'\,d^n\mathbf{x}'.
\end{equation}
For a pulsed line source oriented along the $x_1$ line given by $x_2=0$ line, $S(\mathbf{x},t)=\delta(x_2)\delta(t)$, hence the concentration field evolves as the Gaussian field
\begin{equation}
c(\mathbf{x},t)=\int_{-\infty}^\infty G(x_1,x_2,t|x_1',0,0) dx_1'=\frac{\exp\left(-\frac{x_2^2}{2\sigma(t)^2}\right)}{\sqrt{2\pi}\sigma(t)},\quad \sigma(t)=\sqrt{\frac{2D_m}{\lambda_\infty(1+\coth(\lambda_\infty t))}}.
\end{equation}
Hence the line thickness converges to the Batchelor scale as $\sigma(t)\rightarrow D_m/\lambda_\infty=s_B$.
%Pulsed injection of the material line considered in \S\S~\ref{subsec:lines} generates a line in the $x_2$ direction of initial width $\Sigma_{0,33}(X_2)$ and duration $\Sigma_{0,11}(X_2)$. As concentration gradients are negligible in the $x_2$ direction, only the 1 and 3 components of the covariance matrix play a role
%\begin{equation}
%\boldsymbol\Sigma_\bot(t)=\left(\begin{array}{cc}
%\Sigma_{11}(t) & \Sigma_{13}(t) \\
%\Sigma_{13}(t) & \Sigma_{33}(t)
%\end{array}\right),
%\end{equation}

%%%%%%%%%%%%%%%%%%%%%%%%%%%%%%%%%%%%%%%%%%%%%%%%%%%

\section{Blob Evolution in Steady Flows}

For steady 2D and 3D flows, the $\hat{\mathbf{e}}_1^\prime$ direction of the Protean coordinate frame aligns with the local velocity field, while the $\hat{\mathbf{e}}_2^\prime$ and $\hat{\mathbf{e}}_3^\prime$ directions are transverse to the flow direction. Conversely, for unsteady flows, the Protean coordinate frame ``tumbles'' with no correlation to the underlying velocity field. Due to this alignment, $\epsilon^\prime_{11}(t)=||\mathbf{v}\cdot\nabla\mathbf{v}\cdot\mathbf{v}||/v^2$, and from (\ref{eqn:Fdiag}), the $F_{11}$ component simply fluctuates as per the local velocity magnitude as
\begin{equation}
F^\prime_{11}(t)=\frac{v^\prime(t)}{v^\prime(0)},
\end{equation}
where $v^\prime(t)\equiv||\mathbf{v}^\prime(\mathbf{x}^\prime=\mathbf{0},t)||$. As such, for steady flows, persistent fluid stretching cannot occur in the streamwise direction and associated Lyapunov exponent $\lambda_1=\langle\epsilon^\prime_{11}\rangle$ is zero. This constraint, coupled with (\ref{eqn:lyapzero}) means that, in accordance with the Poincar\'{e}-Bendixson theorem, that steady 2D flows are non-chaotic (i.e. $\lambda_2=0$), and steady 3D flows only admit one unique Lyapunov exponent (where $\lambda_2=-\lambda_3$). These constraints limit evolution of the deformation tensor $\mathbf{F}^\prime(t)$ and the solute blob in 2D and 3D flows.

\subsection{Blob Evolution in Steady 2D Flows}
As 2D steady flows cannot admit chaotic trajectories, then $\lambda_i=0$ for $i=1:2$, and the diagonal components of $\mathbf{F}^\prime$ simply fluctuate with time. For divergence-free flows, these components evolve as $F^\prime_{11}(t)=1/F^\prime_{22}(t)=v^\prime(t)/v^\prime(0)$. Hence the only persistent fluid deformation occurs via shear deformation as characterised by $F_{12}^\prime(t)$, which is driven by the local shear rate $\epsilon^\prime_{12}(t)$. For steady shear flow with shear rate $\epsilon_{12}^\prime(t)=\gamma$, $\mathbf{v}(\mathbf{x})=\hat{\mathbf{e}}_1\gamma x_2$ and the moving $\mathbf{x}^\prime$ and stationary $\mathbf{x}$ frames coincide. In this case the deformation gradient is simply
\begin{align}
\mathbf{F}^\prime(t)=\left(
\begin{array}{cc}
1 & \gamma t\\
0 & 1
\end{array}
\right) &&
\mathbf{F}^\prime(t)^{-1}=\left(
\begin{array}{cc}
1 & -\gamma t\\
0 & 1
\end{array}
\right).
\end{align}
For initial condition $\boldsymbol\Sigma_0=\text{diag}(\sigma_{11},\sigma_{22})$ undergoing molecular diffusion $\mathbb{D}_0=D_m\mathbf{1}$, the reversible and irreversible contributions to the blob covariance then evolve as~\cite{Dentz:2016aa}
\begin{align}
\boldsymbol\Delta(t)=\left(
\begin{array}{cc}
\sigma_{11}+\sigma_{22}\gamma^2 t^2 & \sigma_{22}\gamma t\\
\sigma_{22}\gamma t & \sigma_{22}
\end{array}
\right) &&
\boldsymbol\Lambda(t)=D_m\left(
\begin{array}{cc}
2t+\frac{\gamma^3 t^3}{3} & \gamma t^2\\
\gamma t^2 & 2 t
\end{array}
\right),
\end{align}
where $\gamma t$ terms reflect accelerated dispersion due to linear stretching of the blob under constant shear. For a point-wise injection ($\boldsymbol\Sigma_0=\mathbf{0}$), the dilution index evolves quadratically in time due to shear as
\begin{equation}
E(t)=\frac{2e\pi D_m}{c_0}\sqrt{4 t^2+\frac{1}{3}t^4\gamma^2}.
\end{equation}
For unsteady divergence-free 2D flows in general, the fluctuations along streamlines lead to the deformation gradient tensor
\begin{align}
\mathbf{F}^\prime(t)=\left(
\begin{array}{cc}
\frac{v^\prime(t)}{v^\prime(0)} & r(t)\\
0 & \frac{v^\prime(0)}{v^\prime(t)}
\end{array}
\right) &&
\mathbf{F}^\prime(t)^{-1}=\left(
\begin{array}{cc}
\frac{v^\prime(0)}{v^\prime(t)} & -r(t)\\
0 & \frac{v^\prime(0)}{v^\prime(t)}
\end{array}
\right),
\end{align}
where $r(t)\equiv\frac{v^\prime(t)}{v^\prime(0)}\int_0^t\epsilon_{12}^\prime(t^\prime)\frac{v^\prime(0)^2}{v^\prime(t^\prime)^2}dt^\prime$ quantifies stretching due to shear. \citet{Dentz:2016aa} develop a CTRW for such fluid stretching in steady 2D flows, and show that correlation between shear (characterised by $\epsilon_{12}(t)$) and velocity fluctuations in (\ref{eqn:F12}) generates distinct regimes of fluid stretching behaviour (characterised by $r(t)$), ranging from from sub-diffusive to super-linear: $r(t)\sim t^r$, $0<r<2$. Under molecular diffusion the blob covariance evolves~\citep{Dentz:2016aa} according to the reversbile
\begin{align}
\boldsymbol\Delta(t)=\left(
\begin{array}{cc}
\sigma_{11}\frac{v^\prime(t)^2}{v^\prime(0)^2}+\sigma_{22}r(t)^2 & \sigma_{22}\frac{r(t)v^\prime(0)}{v^\prime(t)}\\
 \sigma_{22}\frac{r(t)v^\prime(0)}{v^\prime(t)} & \sigma_{22}\frac{v^\prime(0)^2}{v^\prime(t)^2},
\end{array}
\right)
\end{align}
and irreversible contributions:
\begin{align}
%\boldsymbol\Lambda(t)=D_m\left(
%\begin{array}{cc}
&\Lambda_{11}(t)/D_m=V_2(t)r(t)^2+\frac{v^\prime(t)^2}{v^\prime(0)^2}(R_2(t)+V_{-2}(t))-2r(t)V_R(t)\frac{v^\prime(t)}{v^\prime(0)},\\
&\Lambda_{12}(t)/D_m=\Lambda_{21}(t)/D_m=\frac{v^\prime(0)}{v^\prime(t)}r(t)V_2(t)-V_R(t),\\
&\Lambda_{22}(t)/D_m=\frac{v^\prime(t)^2}{v^\prime(0)^2}V_{-2}(t),
%\end{array}
%\right),
\end{align}
where $V_i(t)\equiv\int_0^t \frac{v^\prime(t^\prime)^i}{v^\prime(0)^i}dt^\prime$, $R_2(t)\equiv\int_0^t r(t^\prime)^2dt^\prime$, $V_R(t)\equiv.\int_0^t r(t)\frac{v^\prime(t^\prime)}{v^\prime(0)}dt^\prime$. The dilution index for a point-wise injection is then
\begin{equation}
E(t)=\frac{4e\pi D_m}{c_0}\sqrt{(R_2(t)+V_{-2}(t))V_2(t)-V_R(t)^2}.
\end{equation}
As $V_i(t)\sim t$, $V_R(t)\sim t^{r+1}$ and $R_2(t)\sim t^{2r+1}$, at long times this scales as
\begin{equation}
E(t)\rightarrow\frac{4e\pi D_m}{c_0}\sqrt{R_2(t)V_2(t)-V_R(t)^2}\sim t^{r+1}.
\end{equation}
\citep{Dentz:2018aa} show that for mildly heterogeneous porous media (characterised by log-conductivity variance $\sigma_f<1$), the dilution index grows as $E(t)\sim t^{3/2}$ due to correlated fluctuations in the local velocity and shear rate.

%%%%%%%%%%%%%%%%%%%%%%%%%%%%%%%%%%%%%%%%%%%%%%%%
\subsection{Blob Evolution in Steady Integrable 3D Flows}

In general, steady 3D flows can admit Lagrangian chaos, exponential stretching of material elements and a positive Lyapunov exponent $\lambda_\infty$. However, an important class of steady 3D flows are \emph{integrable} flows, which contains several constants of motion (invariants) that constrain the Lagrangian kinematics and fluid deformation, precluding exponential stretching of material elements. One important such flow is steady Darcy flow in a porous medium with a smooth, isotropic conductivity field $k(\mathbf{x})$, as is described by the Darcy equation
\begin{equation}
\mathbf{v}(\mathbf{x})=-k(\mathbf{x})\nabla\phi,
\end{equation}
where $\phi$ is the flow potential. This class of flow has helicity density (defined as $h(\mathbf{x})=\mathbf{v}(\mathbf{x})\cdot\nabla\times\mathbf{v}(\mathbf{x})$) that is identically zero everywhere ($h(\mathbf{x})=0$), and can be shown~\cite{Yoshida:2009aa} to admit two invariants $\psi_1(\mathbf{x})$, $\psi_2(\mathbf{x})$ that acts as streamfunctions of the flow~\citep{Lester:2022aa} as
\begin{equation}
\mathbf{v}(\mathbf{x})=\nabla\psi_1(\mathbf{x})\times\nabla \psi_2(\mathbf{x})=-k(\mathbf{x})\nabla\phi,\label{eqn:sfunc}
\end{equation}
where streamlines arise from the intersection of members of each streamfunction family, and invariance of these streamlines with respect to these streamfunctions is given from (\ref{eqn:sfunc}) as $\mathbf{v}(\mathbf{x})\cdot\nabla\phi_i(\mathbf{x})=0, i=1,2$. The existence of these invariants renders the flow to be integrable, and hence non-chaotic~\citep{Arnold:1965aa}, hence all of the Lyapunov exponents of the flow are zero: $\lambda_i=0, i=1:3.$ In the Protean frame, the helicity-free condition leads to the result that the shear component $\epsilon_{23}^\prime(t)$ transverse to the streamline direction must be zero
\begin{equation}
h(\mathbf{x})=\mathbf{v}^\prime(\mathbf{x}^\prime)\cdot\nabla\times\mathbf{v^\prime}(\mathbf{x}^\prime)=v_i^\prime\epsilon_{ijk}v^\prime_{j,k}=0=v_1^\prime(\epsilon^\prime_{23}-\epsilon^\prime_{32})=v_1^\prime\epsilon^\prime_{23}=0,
\end{equation}
and so the steady 3D flow decouples into a pair of superposed 2D flows. Due to these constraints the deformation gradient tensor is simply 
\begin{align}
\mathbf{F}^\prime(t)=\left(
\begin{array}{ccc}
\frac{v^\prime(t)}{v^\prime(0)} & r_2(t) & r_3(t)\\
0 & \sqrt{\frac{v^\prime(0)}{v^\prime(t)}}\frac{m(t)}{m(0)} & 0\\
0 & 0 & \sqrt{\frac{v^\prime(0)}{v^\prime(t)}}\frac{m(0)}{m(t)}
\end{array}
\right)
\end{align}
\begin{align}
\mathbf{F}^\prime(t)^{-1}=\left(
\begin{array}{ccc}
\frac{v^\prime(0)}{v^\prime(t)} & -r_2(t)\sqrt{\frac{v^\prime(0)}{v^\prime(t)}}\frac{m(0)}{m(t)} & -r_3(t)\sqrt{\frac{v^\prime(0)}{v^\prime(t)}}\frac{m(t)}{m(0)}\\
0 & \sqrt{\frac{v^\prime(t)}{v^\prime(0)}}\frac{m(0)}{m(t)} & 0\\
0 & 0 & \sqrt{\frac{v^\prime(t)}{v^\prime(0)}}\frac{m(t)}{m(0)}
\end{array}
\right),
\end{align}
where $m(t)$ is a solely fluctuating function that characterises the transient relative stretching between the $\hat{\mathbf{e}}_2$ and $\hat{\mathbf{e}}_3$ directions. Similar to the steady 2D flow, for steady integrable 3D flow the shear deformations
\begin{align}
r_2(t)=&\frac{v^\prime(t)}{v^\prime(0)}\int_0^t \epsilon_{12}^\prime(t^\prime)\left(\frac{v^\prime(0)}{v^\prime(t^\prime)}\right)^{3/2}\frac{m(t^\prime)}{m(0)}dt^\prime\\
 r_3(t)=&\frac{v^\prime(t)}{v^\prime(0)}\int_0^t \epsilon_{13}^\prime(t^\prime)\left(\frac{v^\prime(0)}{v^\prime(t^\prime)}\right)^{3/2}\frac{m(0)}{m(t^\prime)}dt^\prime
\end{align}
also both grow as $r(t)\sim t^r$ with $0<r<2$. From the deformation tensor, the reversible contribution to blob covariance is then
\begin{align}
&\Delta_{11}(t)=\sigma_{11}\frac{v^\prime(t)^2}{v^\prime(0)^2}+\sigma_{22}r_2(t)^2+\sigma_{33}r_3(t)^2,\\
&\Delta_{22}(t)=\sigma_{22}r_2(t)\sqrt{\frac{v^\prime(0)}{v^\prime(t)}}\frac{m(t)}{m(0)},\\
&\Delta_{33}(t)=\sigma_{22}\frac{v^\prime(0)}{v^\prime(t)}\frac{m(t)^2}{m(0)^2},\\
&\Delta_{12}(t)=\Delta_{21}(t)=\sigma_{22}r_2(t)\sqrt{\frac{v^\prime(0)}{v^\prime(t)}}\frac{m(t)}{m(0)},\\
&\Delta_{13}(t)=\Delta_{31}(t)=\sigma_{33}r_3(t)\sqrt{\frac{v^\prime(0)}{v^\prime(t)}}\frac{m(0)}{m(t)}, \\
&\Delta_{23}(t)=\Delta_{32}(t)=0,
\end{align}
whereas the irreversible contribution evolves as
\begin{equation}
\begin{split}
\Lambda_{11}(t)/D_m=&I^2_{1,-2,0}r_2(t)^2+I^3_{1,2,0}r_3(t)^2-2\frac{v^\prime(t)}{v^\prime(0)}\left(I^2_{0,-2,1}r_2(t)+I^3_{0,2,1}r_3(t)\right)\\
+&\frac{v^\prime(t)^2}{v^\prime(0)^2}\left(V_{-2}+I_{-1,-2,2}+I_{-1,2,3}\right),
\end{split}
\end{equation}
\begin{align}
&\Lambda_{22}(t)/D_m=\frac{m(t)^2}{m(0)^2}\frac{v(0)}{v(t)}I^2_{1,-2,0},\\
&\Lambda_{33}(t)/D_m=\frac{m(0)^2}{m(t)^2}\frac{v(0)}{v(t)}I^3_{1,2,0},\\
&\Lambda_{12}(t)/D_m=\Lambda_{21}(t)/D_m=\frac{m(t)}{m(0)}\sqrt{\frac{v(0)}{v(t)}}I^2_{1,-2,0}r_2(t)-\frac{m(t)}{m(0)}\sqrt{\frac{v(t)}{v(0)}}I^2_{0,-2,1},\\
&\Lambda_{13}(t)/D_m=\Lambda_{31}(t)/D_m=\frac{m(0)}{m(t)}\sqrt{\frac{v(0)}{v(t)}}I^3_{1,2,0}r_3(t)-\frac{m(0)}{m(t)}\sqrt{\frac{v(t)}{v(0)}}I^3_{0,2,1},\\
&\Lambda_{23}(t)/D_m=\Lambda_{32}(t)/D_m=0,
\end{align}
where 
$I^p_{i,j,k}(t)\equiv\int_0^t \frac{v^\prime(t^\prime)^i}{v^\prime(0)^i}\frac{m(t^\prime)^j}{m(0)^j}r_p(t^\prime)^kdt^\prime$. The dilution index for point-wise injection is then the 3D analogue of the 2D result
\begin{equation}
E(t)=\frac{4\sqrt{2}e\pi D_m}{c_0}\sqrt{-I_{1,-2}(I^3_{0,2,1})^2-I_{1,2}(I^2_{0,-2,1})^2+I_{1,-2}I_{1,2}\left(V_{-2}+I^2_{-1,-2,2}+I^3_{-1,2,2}\right)},
\end{equation}
where in the limit of large $t$, the $I^2_{-1,-2,2}$ and $I^3_{-1,2,2}$ terms dominate as they scale as $t^{2r}$, while the $I_{1,-2}$ and $I_{1,2}$ terms scale as $t$, hence the same scaling as () is recovered
\begin{equation}
E(t)\rightarrow\frac{4\sqrt{2}e\pi D_m}{c_0}\sqrt{I_{1,-2}I_{1,2}\left(V_{-2}+I^2_{-1,-2,2}+I^3_{-1,2,2}\right)}\sim t^{1+r}.
\end{equation}

%%%%%%%%%%%%%%%%%%%%%%%%%%%%%%%%%%%%%%%%%%%%%%%%
\subsection{Blob Evolution in Steady Chaotic 3D Flows}

For steady chaotic 3D flows, the Protean deformation gradient tensor and its inverse may be written in the most general form as
\begin{align}
\mathbf{F}^\prime(t)=\left(
\begin{array}{ccc}
\frac{v^\prime(t)}{v^\prime(0)} & \gamma_{12}(t)e^{\mu(t)} & \gamma_{13}(t)e^{\mu(t)}\\
0 & \sqrt{\frac{v^\prime(0)}{v^\prime(t)}} e^{\mu(t)} & \gamma_{23}(t)e^{\mu(t)}\\
0 & 0 & \sqrt{\frac{v^\prime(0)}{v^\prime(t)}} e^{-\mu(t)},
\end{array}
\right)
\end{align}

\begin{align}
\mathbf{F}^\prime(t)^{-1}=
\left(
\begin{array}{ccc}
\frac{v^\prime(0)}{v^\prime(t)} 
& -\gamma_{12}(t)\sqrt{\frac{v^\prime(0)}{v^\prime(t)}} 
& e^{\mu(t)}\left(\gamma_{12}(t)\gamma_{23}(t)-\gamma_{13}(t)\sqrt{\frac{v^\prime(0)}{v^\prime(t)}}\right)
\\
0 
& \sqrt{\frac{v^\prime(t)}{v^\prime(0)}} e^{-\mu(t)} 
& -\gamma_{23}(t)\frac{v^\prime(t)}{v^\prime(0)}e^{\mu(t)} 
\\
0 
& 0
& \sqrt{\frac{v^\prime(t)}{v^\prime(0)}} e^{\mu(t)},
\end{array}
\right)
\end{align}
where the 2 direction is aligned with net fluid stretching, and the 3 direction is aligned with net fluid contraction. Hence the Lyapunov exponent of this flow is $\lambda=\langle\mu(t)\rangle$. Note there is no net stretching in the 1-drection due to the steady nature of the flow. The shears $\gamma_{ij}$ operate in the two longitudinal planes (12, 13) and the transverse plane ( 23) where $\gamma_{23}$ in the Protean frame is directly associated with the helicity as $h=\mathbf{v}\cdot(\boldsymbol\epsilon_{ijk}:\nabla\mathbf{v})=v(\epsilon_{23}=v\gamma_{23}$. The expressions for $\boldsymbol\Delta(t)$ and $\boldsymbol\Lambda(t)$ are too complex to be compactly shown here, but the dilution index $E(t)$ for this flow then evolves as

\begin{equation}
E(t)=\frac{4\sqrt{2}e\pi D_m}{c_0}\sqrt{-I_{1,-2}(I^3_{0,2,1})^2-I_{1,2}(I^2_{0,-2,1})^2+I_{1,-2}I_{1,2}\left(V_{-2}+I^2_{-1,-2,2}+I^3_{-1,2,2}\right)},
\end{equation}
where in the limit of large $t$, the $I^2_{-1,-2,2}$ and $I^3_{-1,2,2}$ terms dominate as they scale as $t^{2r}$, while the $I_{1,-2}$ and $I_{1,2}$ terms scale as $t$, hence the same scaling as () is recovered
\begin{equation}
E(t)\rightarrow\frac{4\sqrt{2}e\pi D_m}{c_0}\sqrt{I_{1,-2}I_{1,2}\left(V_{-2}+I^2_{-1,-2,2}+I^3_{-1,2,2}\right)}\sim t^{1+r}.
\end{equation}

%%%%%%%%%%%%%%%%%%%%%%%%%%%%%%%%%%%%%%%%%%%%%%%%%%%
\section{Conclusions}
%%%%%%%%%%%%%%%%%%%%%%%%%%%%%%%%%%%%%%%%%%%%%%%%%%%

These results provide a basis for understanding and quantifying solute mixing in a wide range of flows from 2D to 3D, steady to unsteady, regular to chaotic, Stokes to turbulent subject to a range of injection conditions from continuous to pulsed, point-wise to line injection. We derive a Green's function for the advection diffusion equation based on linearisation of the velocity field, and show that the mixing of Gaussian solute plumes is governed solely by the deformation history of fluid elements Significant simplifications arise when the deformation tensor is placed in the Protean~\cite{Lester:2018aa} coordinate frame (which aligns with the maximum and minimum fluid stretching directions and renders the deformation tensor upper triangular), allowing evolution of the Gaussian covariance matrix to be expressed in physically relevant terms such as Lyapunov exponents, longitudinal and transverse shear rates. We couch these results in terms of evolution of the maximum concentration and scalar entropy of Gaussian plume for a range of flows and injection protocols. These results provide significant insights into the processes that govern mixing in a broad range of flows and provide the building blocks of efficient numerical methods using e.g. radial basis functions for solute transport and mixing in arbitrary flows.

\bibliography{database.ref}
\bibliographystyle{jfm}

\end{document}